\documentclass{iaus}
\usepackage{graphicx}

\title[~~Consequences of the {\it HST} proper motions] 
{The Binarity of the Clouds and the \\ Formation of the Magellanic Stream}

\author[Besla et al]   
{Gurtina Besla$^1$, Nitya Kallivayalil$^2$, Lars Hernquist$^1$, Roeland P. van der Marel$^3$, T.J. Cox$^1$, Brant Robertson$^4$ \and Charles Alcock$^1$}

\affiliation{$^1$Harvard Smithsonian Center for Astrophysics, 60 Garden St.,
Cambridge, MA, 02138, USA \\ email: {\tt gbesla@cfa.harvard.edu} \\[\affilskip]
$^2$Pappalardo Fellow, MIT Kavli Institute for Astrophysics and Space Research,\\ 70 Vassar St., 37-66H, Cambridge, MA, 02139, USA \\[\affilskip]$^3$Space Telescope Science Institute, 3700 San Martin Dr., Baltimore, MD, 21218, USA \\[\affilskip]$^4$Spitzer Fellow; Kavli Institute for Cosmological Physics and Dept of Astronomy and Astrophysics, University of Chicago, 933 East 56th St., Chicago, IL, 60637, USA}

\pubyear{2008}
\volume{256}  
\pagerange{119--126}
\jname{The Magellanic System: Stars, Gas, and Galaxies}
\editors{Jacco Th. van Loon \& Joana M. Oliveira, eds.}
\begin{document}

\maketitle

\begin{abstract}
The {\it HST} proper motion (PM) measurements of the Clouds have severe implications for their interaction history with the Milky Way (MW) and with each other. The Clouds are likely on their first passage about the MW and the SMC's orbit about the LMC is better described as quasi-periodic rather than circular. Binary L/SMC orbits that satisfy observational constraints on their mutual interaction history (e.g. the formation of the Magellanic Bridge during a collision between the Clouds $\sim$300 Myr ago) can be located within 1$\sigma$ of the mean PMs. However, these binary orbits are not co-located with the Magellanic Stream (MS) when projected on the plane of the sky and the line-of-sight velocity gradient along the LMC's orbit is significantly steeper than that along the MS.  These combined results ultimately rule out a purely tidal origin for the MS: tides are ineffective without multiple pericentric passages and can neither decrease the velocity gradient nor explain the offset stream in a polar orbit configuration. Alternatively, ram pressure stripping of an extended gaseous disk may naturally explain the deviation. The offset also suggests that observations of the little-explored region between RA 21$^\mathrm{h}$ and 23$^\mathrm{h}$ are crucial for characterizing the full extent of the MS.
\keywords{galaxies: Magellanic Clouds, galaxies: kinematics and dynamics, galaxies: interactions}
\end{abstract}

\firstsection
\section{Introduction}

The recent high-precision proper motion (PM) measurements of the L/SMC determined by \cite[Kallivayalil et al. (2006a, 2006b - hereafter K06a and K06b; see also these proceedings)]{K06a} imply that the Magellanic Clouds are moving $\sim$100 km/s faster than previously estimated and now approach the escape velocity of the Milky Way (MW). \cite[Besla et al. (2007)]{B07} (hereafter B07) re-examined the orbital history of the Clouds using the new PMs and a $\Lambda$CDM-motivated MW model and found that the L/SMC are either on their first passage about the MW or, if the mass of the MW is $> 2\times10^{12}M_\odot$, that their orbital period and apogalacticon distance are a factor of three larger than previously estimated. This means that models of the Magellanic Stream (MS) need to reconcile the fact that although the efficient removal of material via tides and/or ram pressure requires multiple pericentric passages through regions of high gas density, the PMs imply that the Clouds did not pass through perigalacticon during the past $\ge$5 Gyr (this is true even if a high mass MW model is adopted). While the most dramatic consequence of the new PMs is the limit they place on the interaction timescale of the Clouds with the MW, there are a number of other equally disconcerting implications: the relative velocity between the Clouds has increased such that only a small fraction of the orbits within the PM error space allow for stable binary L/SMC orbits (K06b, B07); the velocity gradient along the orbit is much steeper than that observed along the MS; and the past orbits are not co-located with the MS on the plane of the sky (B07). In these proceedings the listed factors are further explored and used to argue that the MS is not a tidal tail.

\section{Do the Clouds form a binary system?}

\begin{figure}[b]
  \begin{center}
    \includegraphics[width=3.2in]{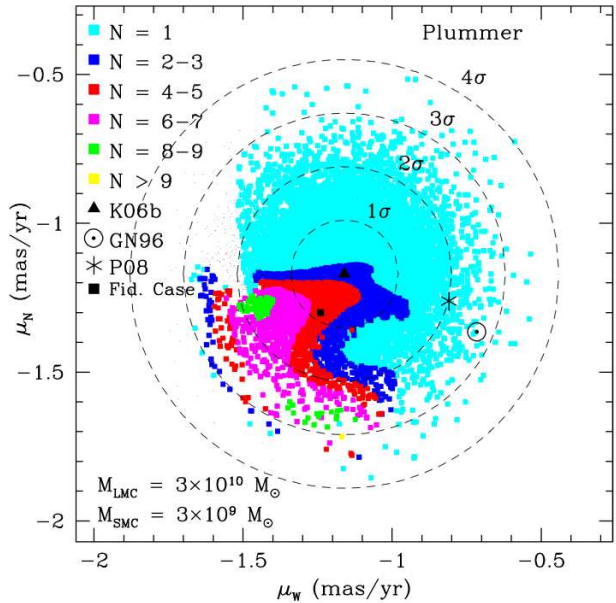} 
    \caption{Ten thousand points randomly sampled from the (4 $\sigma$) K06b PM error space for the SMC (where the mean value is indicated by the triangle). Each corresponds to a unique 3D velocity vector and is color coded by the number of times the separation between the Clouds reaches a minimum within a Hubble time. The circled dot indicates the GN96 PM for the SMC and the asterisk corresponds to the mean of the \cite[Piatek et al. (2008)]{P08} (P08) re-analysis of the K06b data - neither correspond to long-lived binary states. The Clouds are modeled as Plummer potentials with masses of $M_{LMC}=3\times10^{10}M_\odot$ and $M_{SMC}=3\times 10^9 M_\odot$ and the MW is modeled as a NFW halo with a total mass of $10^{12}M_\odot$ as described in B07. The LMC is assumed to be moving with the mean K06a PM (v=378 km/s). The black square represents a solution for the SMC's PM that allows for close passages between the Clouds at characteristic timescales (see Fig.\ \ref{fig2}) and is our fiducial case.}
    \label{fig1}
  \end{center}
\end{figure}

\begin{figure}[t]
  \begin{center}
    \includegraphics[width=3.2in]{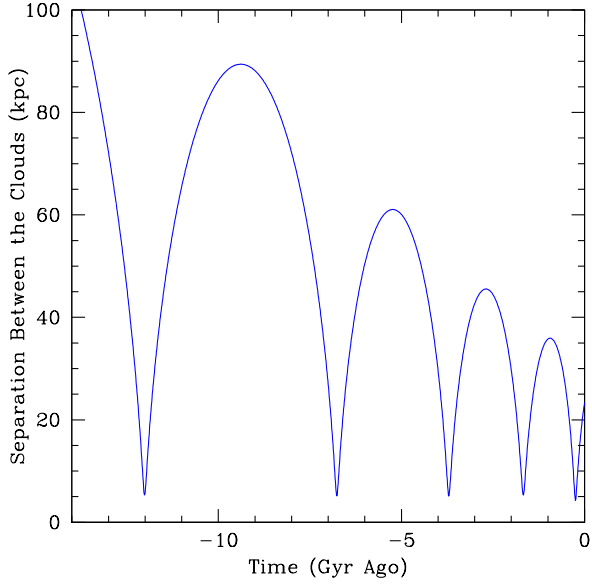} 
    \caption{The separation between the Clouds is plotted as a function of time in the past for the fiducial SMC orbit indicated by the black square in Fig.\ \ref{fig1} and assuming a mass ratio of 10:1 between the L/SMC. The separation reaches a minimum at $\sim$300 Myr and $\sim$1.5 Gyr in the past, corresponding to the formation times for the bridge and the MS. }
    \label{fig2}
  \end{center}
\end{figure}

Doubt concerning the binarity of the Clouds is particularly troubling, as a recent chance encounter between dwarf galaxies in the MW's halo is improbable if they did not have a common origin. To address this issue, ten thousand points were randomly drawn from the SMC PM error space (K06b), each corresponding to a unique velocity vector and orbit (Fig.\ \ref{fig1}). Bound orbits are identified and color coded based on the number of times the separation between the Clouds reaches a minimum, assuming a mass ratio of 10:1 between the L/SMC (although the mass ratio is not well constrained). 
Orbits with only one close encounter (like for the SMC PM determined in the re-analysis of the K06b data by \cite[Piatek et al. 2008]{P08}, hereafter P08) are not stable binary systems.  The new LMC PM also implies that orbits where the SMC traces the MS on the plane of the sky (like that chosen by \cite[Gardiner \& Noguchi 1996]{GN96}, hereafter GN96) are no longer binary orbits. It is clear from Fig.\ \ref{fig1} that stable binary orbits exist within 1$\sigma$ of the mean K06b value - however, in all cases the SMC's orbit about the LMC is highly eccentric (Fig.\ \ref{fig2}), which differs markedly from the conventional view that the SMC is in a circular orbit about the LMC (GN96, \cite[Gardiner et al. 1994]{GSF94}). It should also be noted that the likelihood of finding a binary L/SMC system that is stable against the tidal force exerted by the MW decreases if the MW's mass is increased.

We further require that the last close encounter between the Clouds occurred $\sim$300 Myr ago, corresponding to the formation timescale of the Magellanic Bridge (\cite[Harris 2007]{H07}), and that a second close encounter occurs $\sim$1.5 Gyr ago, a timeframe conventionally adopted for the formation of the MS (GN96). A case that also satisfies these constraints is indicated in Fig.\ \ref{fig1} by the black square and will be referred to as our fiducial SMC orbit. The corresponding orbital evolution of the SMC about the LMC is plotted in Fig.\ \ref{fig2}: the new PMs are not in conflict with known observational constraints on the mutual interaction history of the Clouds. This provides an important consistency check for the K06a,b PMs: if the measurements suffered from some unknown systematics, it would be unlikely for binary orbits to exist within the error space.

\section{Consequences for the Magellanic Stream}

\begin{figure}[t]
  \begin{center}
    \includegraphics[width=3.2in]{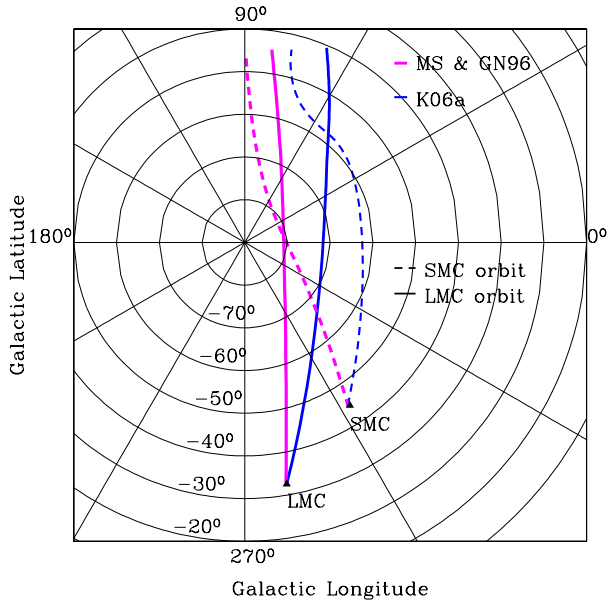} 
    \caption{The past orbits of the LMC (solid lines) and SMC (dashed lines) for our fiducial PMs (blue) and the GN96 PMs (magenta) are mapped as a polar projection in galactic (l,b) coordinates.  The orbits are followed backwards in time from the Clouds' current positions (filled triangles) until they extend 100$^o$ in the sky. The fiducial orbits deviate markedly from the current location of the MS, which is traced by the GN96 orbits.}
    \label{fig3}
  \end{center}
\end{figure}

\begin{figure}[t]
  \begin{center}
    \includegraphics[width=3.2in]{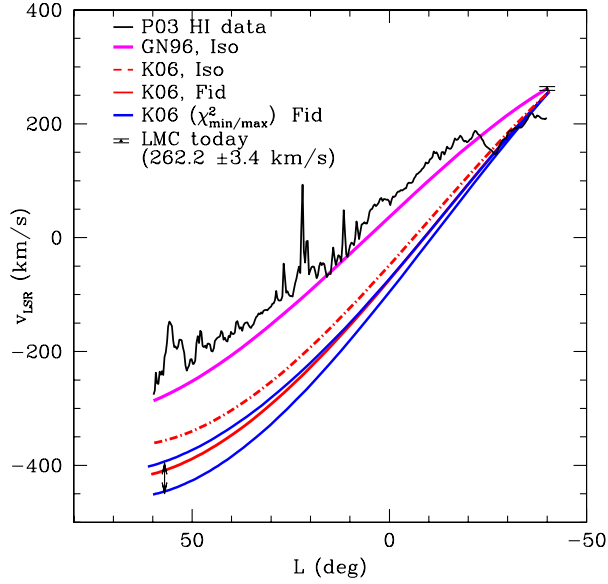} 
    \caption{The line-of-sight velocities (v$_{LSR}$) with respect to the local standard of rest are plotted as a function of Magellanic longitude (L) (see B07, Fig. 20) along the LMC's orbit. The orbital velocity corresponding to the mean PM for the B07 fiducial(isothermal sphere) MW model is indicated by the solid red(dashed red) line. The best and worst $\chi^2$ fits to the MS data within the PM error space are indicated by the arrows and the blue lines. The black line indicates the HI data of the MS from \cite[Putman et al. (2003)]{P03} (P03). The velocity gradients along the new orbits in both the isothermal sphere and fiducial NFW MW models are significantly steeper than the GN96 results (magenta line), which were contrived to trace the velocity data.}  
    \label{fig4}
  \end{center}
\end{figure}

The spatial location of the fiducial orbit on the plane of sky and the line-of-sight velocity gradient along it are compared to the observed properties of the MS. The GN96 orbits were a priori chosen to trace both the spatial location and velocity structure of the MS, but this is an assumption. Indeed, from Fig.\ \ref{fig3}, the LMC's orbit using the new PM is found to be offset from the MS (indicated by the GN96 orbits) by roughly $10^o$.  The offset arises because the north component of the LMC's PM vector as defined by K06a, the re-analysis by P08, {\it and} the weighted average of all PM measurements prior to 2002 (\cite[van der Marel et al. 2002]{vdM02}), is not consistent with 0 (which was the assumption made by GN96): this result is thus independent of the MW model (B07). Furthermore, the SMC must have a similar tangential motion as the LMC in order to maintain a binary state, meaning that our fiducial SMC orbit deviates even further from the MS than that of the LMC.  In addition, the line-of-sight velocity gradient along the LMC's orbit is found to be significantly steeper than that of the MS, reaching velocities $\sim$200 km/s larger than that observed at the same position along the MS (Fig.\ \ref{fig4}).

The offset and steep velocity gradient are unexplainable in a tidal model. While tidal tails may deviate from their progenitor's orbit, they remain confined to the orbital plane (\cite[Choi et al. 2007]{Choi}): since the Clouds are in a polar orbit no deviation is expected in projection in a tidal model. Furthermore, material in tails is accelerated by the gravitational field of the progenitor - however, to explain the observed velocity gradient the opposite needs to occur. Coupling these factors to a first passage scenario strongly suggests that, while tides likely help shape the stream (e.g. the leading arm feature), hydrodynamic processes are the {\it primary} mechanism for the removal of material from the Clouds and for shaping its velocity structure (e.g. via gas drag). 

\begin{figure}[t]
  \begin{center}
    \includegraphics[width=2.6in]{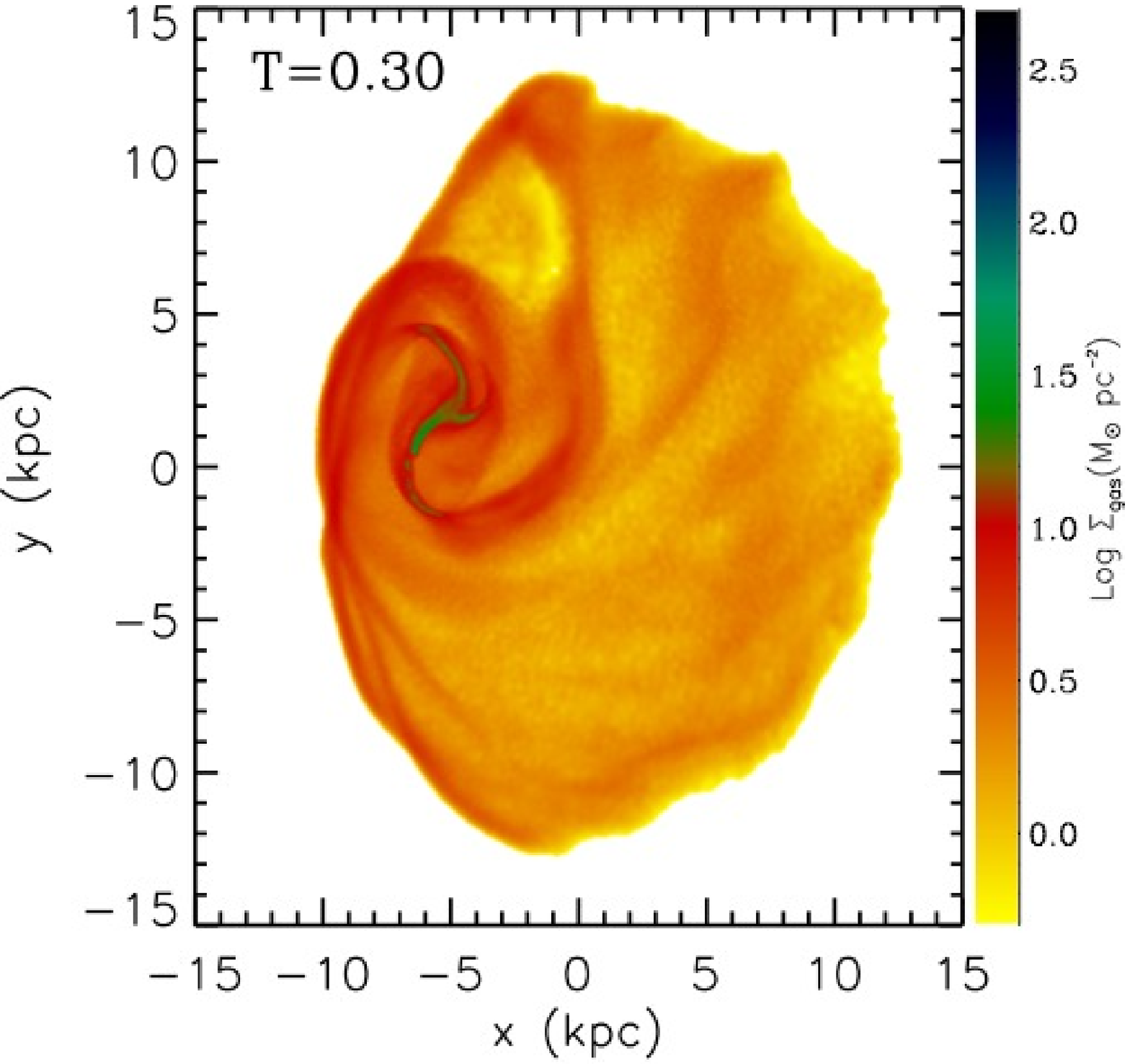}
	\includegraphics[width=2.6in]{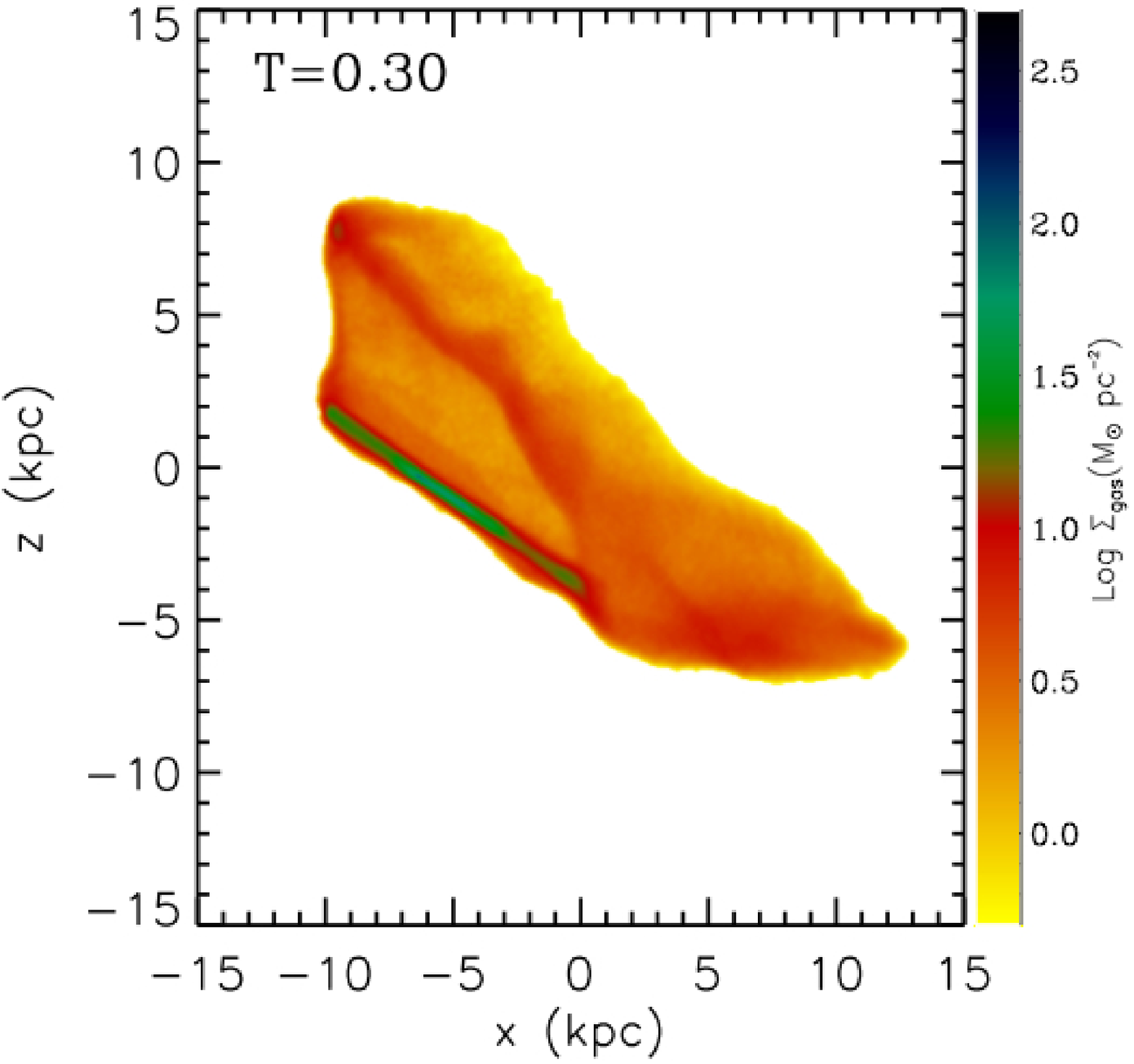}
    \caption{Ram pressure stripping studies of an LMC model with a gaseous disk component that is extended 3x the optical extent and rotating clockwise. The LMC is modeled as an NFW halo with an exponential disk and a total mass of $3\times10^{10}M_\odot$. The left panel is the face-on view and the right is edge-on: in all cases the LMC is moving to the left at 380 km/s and is inclined 30$^o$ relative to its direction of motion. The snapshots indicated the evolution of the gas surface density after 0.3 Gyr. The face-on projection illustrates how material is preferentially removed from the side of the disk rotating in-line with the ram pressure wind. In the edge-on projection, material from the leading edge lags behind that removed from the trailing edge. }  
    \label{fig5}
  \end{center}
\end{figure}

The main difficulty for ram pressure stripping in a first passage scenario is the low halo gas densities at large galactocentric distances. The efficiency of stripping may be improved if material is given an additional kick by stellar feedback (e.g. \cite[Nidever et al 2008]{N08}, \cite[Olano 2004]{Olano}) or if the LMC initially possessed an extended disk of HI like those observed in isolated dIrrs - note that the latter is not a viable initial condition if the LMC were not on its first passage and the former may violate metallicity constraints on the MS which indicate the MS is metal poor (\cite[Gibson et al. 2000]{Gibson}). If ram pressure stripping is efficient, the offset may occur naturally: \cite[Roediger \& Br{\"u}ggen (2006)]{RB06} have shown that material can be removed asymmetrically from gaseous disks that are inclined relative to their line of motion (the LMC's disk is inclined by $30^o$) and caution that tails do not always indicate the direction of motion of the galaxy. These authors considered ram pressure stripping in the context of massive galaxies in cluster environments. We are currently conducting simulations of the formation of the MS via the ram pressure stripping of the Clouds, assuming they initially entered the MW system with extended gaseous disks. 
 Fig. \ref{fig5} illustrates the proposed mechanism at work: here the LMC has been moving at 380 km/s through a box of gas at a uniform temperature of $10^6$ K and density of $10^{-4}$ /cm$^3$ for 300 Myr. Once the material is removed beyond the LMC's tidal radius, the MW's tidal force may then be able to stretch the material to its full extent - but now since the material is removed asymmetrically it will not trace the orbit in projection.

\section{Conclusions}

The new PMs have dramatic implications for phenomenological studies of the Clouds that assume they have undergone multiple pericentric passages about the MW and/or that the SMC is in a circular orbit about the LMC. The orbits deviate spatially from the current location of the MS on the plane of the sky and the velocity gradient along the orbit is much steeper than that observed. These results effectively rule out a purely tidal model for the MS and lend support for hydrodynamical models, such as ram pressure stripping.  The offset further suggests that the Clouds have travelled across the little explored region between RA 21$^\mathrm{h}$ and 23$^\mathrm{h}$ (i.e. the region spanned by the blue lines in Fig.\ \ref{fig3}). \cite[Putman et al. (2003)]{P03} detected diffuse HI in that region that follow similar velocity gradients as the main stream (their Fig. 7), but otherwise material in that region has been largely ignored by observers and theorists alike. The offset orbits suggest that the MS may be significantly more extended than previously believed and further observations along the region of sky they trace are warranted.


\begin{thebibliography}{}

\bibitem[Besla \etal\ (2007)]{B07}
{Besla, G., Kallivayalil, N., Hernquist, L., Robertson, B., Cox, T.J., van der Marel, R.P., \& Alcock, A.} 2007, 
\textit{ApJ}, 668, 949 (B07)

\bibitem[Choi \etal\ (2007)]{Choi}
{Choi,J.-H., Weinberg, M. \& Katz, N.} 2007, 
\textit{MNRAS}, 381, 987

\bibitem[Gardiner \& Noguchi (1996)]{GN96}
{Gardiner, L.T. \& Noguchi, M.} 1996, 
\textit{MNRAS}, 278, 191, (GN96) 

\bibitem[Gardiner \etal\ (1994)]{GSF94}
{Gardiner, L.T., Sawa, T. \& Fujimoto, M.} 1994, 
\textit{MNRAS}, 266, 567 

\bibitem[Gibson \etal\ (2000)]{Gibson}
{Gibson, B.K., Giroux, M.L., Penton, S.V., Putman, M.E., Stocke, J.T., \& Shull, J.M.} 2000, 
\textit{ApJ}, 120, 1830

\bibitem[Harris (2007)]{H07} 
{Harris, J.} 2007, 
\textit{ApJ}, 658, 345

\bibitem[Kallivayalil \etal\ (2006a)]{K06a}
{Kallivayalil, N., van der Marel, R.P., Alcock, C., Axelrod, T., Cook, K.H., Drake, A.J., \& Geha, M.}2006a, 
\textit{ApJ}, 638, 772 (K06a)

\bibitem[Kallivayalil \etal\ (2006b)]{K06b}
{Kallivayalil, N., van der Marel, R.P., \& Alcock, C.}2006b, 
\textit{ApJ}, 652, 1213 (K06b)

\bibitem[Nidever \etal\ (2008)]{N08}
{Nidever, D.L., Majewski, S.R., \& Burton, W.B.} 2008, 
\textit{ApJ}, 679, 432

\bibitem[Olano (2004)]{Olano}
{Olano, C.A.} 2004
\textit{A\&A}, 423, 895

\bibitem[Piatek \etal\ (2008)]{P08}
{Piatek, S., Pryor, C., \& Olszewski, E.W.} 2008,
\textit{ApJ}, 135, 1024 (P08)

\bibitem[Putman \etal\ (2003)]{P03}
{Putman, M.E., Staveley-Smith, L., Freeman, K.C., Gibson, B.K., \& Barnes, D.G.} 2003, 
\textit{ApJ}, 586, 170

\bibitem[Roediger \& Br{\"u}ggen (2006)]{RB06}
{Roediger, E. \& Br{\"u}ggen, M.} 2006,
\textit{MNRAS}, 369, 567

\bibitem[van der Marel \etal\ (2002)]{vdM02}
{van der Marel, R.P., Alves, D.R., Hardy, E., \* Suntzeff, N.B.} 2002,
\textit{AJ}, 124, 2639


\end{thebibliography}
\end{document}